\documentclass[aps,prl,floats,floatfix,twoside,preprintnumbers,superscriptaddress,nofootinbib]{revtex4}

\usepackage{graphicx}

\usepackage{subfig}

\begin{document}

\title{The competitiveness versus the wealth of a country}

\author{Boris~Podobnik}
\affiliation{Center for Polymer Studies and Department of Physics, Boston 
University, Boston, MA 02215, USA}
\affiliation{Zagreb School of Economics and Management,  10 000 Zagreb,
 Croatia}
\affiliation{Faculty of Civil Engineering, University of Rijeka, 51000 
Rijeka, Croatia}

\author{Davor Horvati\'c}
\affiliation{Physics Department, Faculty of Science, University of Zagreb, 
10000 Zagreb, Croatia}

\author{ Dror Y. Kenett}
\affiliation{Center for Polymer Studies and Department of Physics, Boston 
University, Boston, MA 02215, USA}

\author{H.~Eugene~Stanley}
\affiliation{Center for Polymer Studies and Department of Physics, Boston 
University, Boston, MA 02215, USA}

\begin{abstract}

Politicians world-wide frequently promise a better life for their citizens. We find that the probability that a country will increase its {\it per capita} GDP ({\it gdp}) rank within a decade follows an exponential distribution with decay constant $\lambda
= 0.12$. We use the Corruption Perceptions Index (CPI) and the Global Competitiveness Index (GCI) and find that the distribution of change in CPI (GCI) rank follows exponential functions with approximately the same exponent as $\lambda$, suggesting that the dynamics of {\it gdp}, CPI, and GCI may share the same origin. Using the GCI, we develop a new measure, which we call relative competitiveness, to evaluate an economy's competitiveness relative to its {\it gdp}. For all European and EU countries during the 2008-2011 economic downturn we find that the drop in {\it gdp} in more competitive countries relative to {\it gdp} was
substantially smaller than in relatively less competitive countries, which is valuable information for policymakers.

\end{abstract}

\maketitle

An economy's competitiveness is defined as the set of institutions,
policies, and factors that determine the economy's level of productivity
\cite{Schwab}. The level of productivity determines the rates of return
obtained by investments in the economy, which in turn determines the
economy's growth rate
\cite{Schwab,Schultz,Romer2,Lucas,Fischer,Alesina,martin96,barrosalabook}.

Why does competitiveness in some nations increase more rapidly than in
other nations?  Why do some nations seem unconcerned about improving
their competitiveness, even though increasing competitiveness attracts
investment thus increases wealth?  Whether poor countries will be able
to catch up to rich countries has been a question for many years, and
the varying conclusions  drawn  have depended on the approach taken, e.g.,
how many countries are involved or whether size dependence is taken into
consideration
\cite{Solow,Quah93,barrosalabook,martin96,Durlauf96,Sachs,martin06}.
Here we ask what is the probability that a poor country can become a
rich country within a given time period, e.g., within a decade? How are
the dynamics of competitiveness and the dynamics of a country's wealth
related?

We focus on both static and dynamic aspects of competitiveness
\cite{Schwab} by examining the per capita Gross Domestic Product (GDP)
\cite{imf}, denoted as {\it gdp}, and the Global Competitiveness Index
(GCI) (see Methods), which quantifies the institutions, policies, and
factors that control economic prosperity \cite{sala04,Schwab}.  First we
quantify the relationship between the {\it gdp} and the GCI.  Then we
introduce a new measure to assess the competitiveness relative to {\it
  gdp} as the difference ${\cal D}$ between the actual GCI value and the
expected value of GCI obtained from the power-law fit between GCI and
{\it gdp}---the more negative is ${\cal D}$, the smaller will be the
relative competitiveness of a given country.  We examine how the level
of competitiveness affects the dynamics of a country's wealth during a
recession, finding that during the 2008--2011 period EU countries with
positive ${\cal D}$ values experienced a significantly smaller drop in
{\it gdp} than countries with negative ${\cal D}$ values.  The
probability that a country will increase its wealth---quantified by its
rank within a decade---follows an exponential distribution. We relate
this probability to the probability of change in GCI rank, and our
results are consistent with the interesting possibility that the
dynamics of {\it gdp} and the dynamics of GCI may share the same origin.

\section{Results}

\textbf{Competitiveness versus growth.}  Over the past decade, countries
such as Switzerland, Singapore, the Nordic countries, and the USA have
been considered to be the most competitive.  Generally speaking, rich
countries are considered to be more competitive than poor countries,
implying that there is a functional dependence between GCI and {\it
  gdp}.  Here we address two questions: (i) What is the expected level
of competitiveness for a country with a given level of wealth? (ii) What
is the probability that a country will substantially improve its wealth
and its level of competitiveness?

\begin{figure}[!ht]
\centering  
\includegraphics[width=0.4\textwidth]{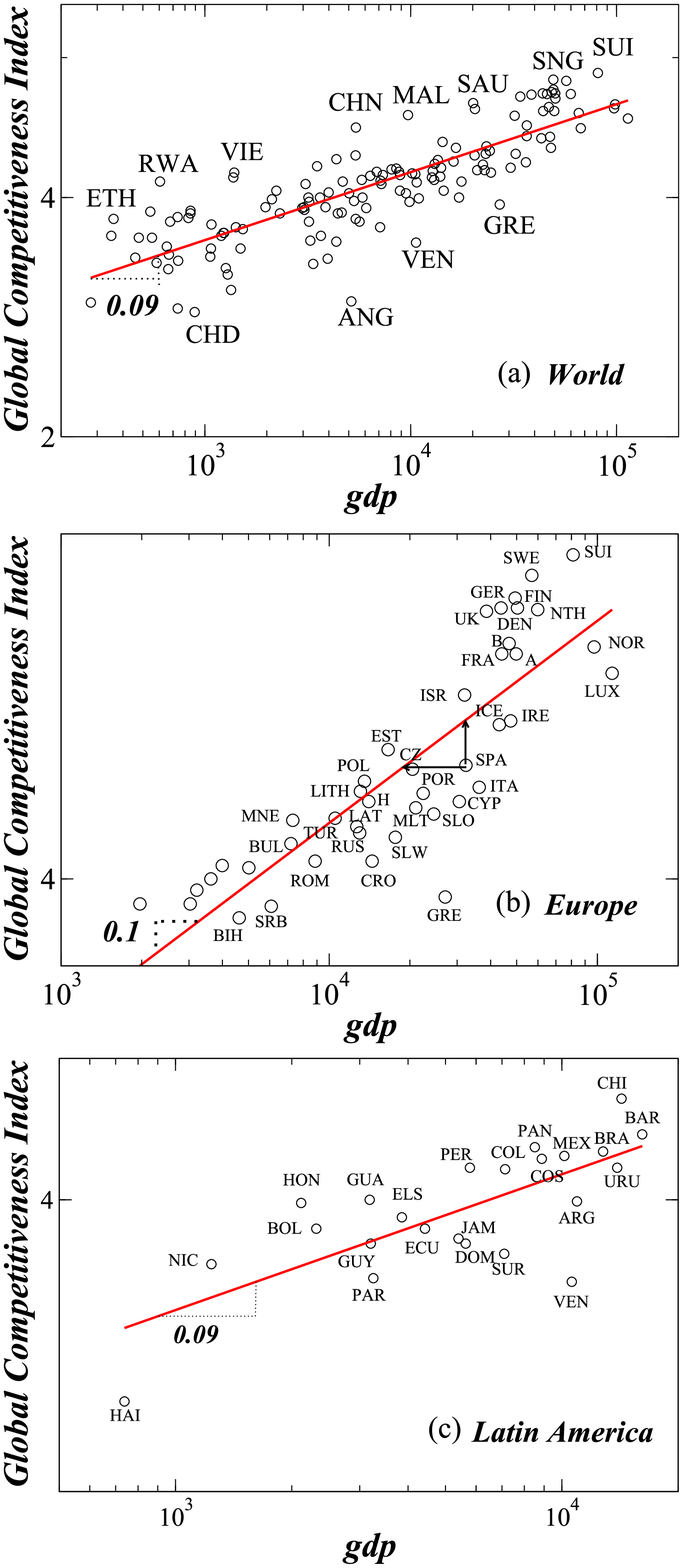}  
\caption{Relative competitiveness measured by GCI versus GDP per capita
  ({\it gdp}) for (a) World, (b) European and (c) Latin American
  countries. Above (below) the power-law fitting line, the level of
  competitiveness is more (less) than we expect for the given country
  wealth, quantified by {\it gdp}. In (b)  Israel is shown, but does not
  contribute to the fitting line. 
  }
\label{GCI}
\end{figure}

Addressing question (i), Figs.~\ref{GCI}(a) and \ref{GCI}(b) show plots
of GCI versus {\it gdp}, both worldwide and for different European
countries. We find a positive functional dependence between GCI and {\it
  gdp}, which we fit with a power law,
\begin{equation}
{\rm GCI}\propto ({\rm {\it gdp}})^{\alpha}
\label{Eq1} 
\end{equation}
with exponent $\alpha \approx 0.1$. Figure~\ref{GCI}(b) shows that
Germany is more competitive than Estonia, but that is to be expected
since the German {\it gdp\/} is much larger than the Estonian {\it gdp}.
But is Germany more competitive with its peer countries---those with a
comparable {\it gdp}---than Estonia with its peer countries? When
comparing competitiveness relative to {\it gdp}, how do we estimate
which countries are better performers and which are worse? In addressing
this question, note that the power-law dependence in Fig.~\ref{GCI}(b)
indicates the expected level of competitiveness for a given level of
country wealth. When a country appears above the fitting line, its level
of competitiveness relative to {\it gdp\/} is greater than expected for a
country with its given wealth level, quantified by {\it gdp}.  In
contrast, countries below the fitting line are less competitive than
expected for a country with its given wealth level. This information is
valuable for investors who need to assess which countries with
comparable {\it gdp\/} values to choose for investment
opportunities. For example, Fig.~\ref{GCI}(b) shows that Poland is more
competitive than Croatia, which we would expect since these two
countries have similar {\it gdp}, but the Polish GCI is higher than the
Croatian GCI.

The regression we obtain also allows us to compare the relative levels
of competitiveness between two countries that belong to two different
wealth brackets.  To this end, we introduce a measure to assess relative
competitiveness
\begin{equation}
{\cal D}\equiv \ln({\rm GCI}) - \langle \ln({\rm GCI}) \rangle ,
 \label{D}
\end{equation}  
i.e., the difference between the actual GCI value and the expected value
of GCI obtained from the power-law fitting line.  For countries below
the power-law regression line, the more negative the difference ${\cal
  D}$, the smaller the relative competitiveness.  Figure~\ref{GCI}(b)
shows that Greece is the least competitive in relative terms among all
the European countries, since Greece has the most negative ${\cal D}$.
Note that Norway seems to be an outlier among the Nordic countries, but
Norway is a huge exporter of oil and thus perhaps does not need to be
competitive.  Thus for a country below the regression line (see Spain),
Fig.~\ref{GCI} suggests (a) the expected level of competitiveness a
country should aspire to in order to achieve at least the average
relative competitiveness (the vertical line), and (b) where the country
may end up if it does not improve its competitiveness (the horizontal
line toward the left).
    
For Latin American countries Fig.~\ref{GCI}(c) shows a power law for the
GCI versus {\it gdp\/} with an exponent $\alpha = 0.09 \pm 0.02$, a value
virtually identical to $\lambda$ found for the entire world and for the
European countries, implying universality in the regression of
Eq.~(\ref{Eq1}).  Note that the country with the lowest GCI value,
Venezuela, is a huge exporter of oil.
  
Up to this point no temporal dependence has been included (year 2011),
and we have used a ``static'' approach in assessing levels of
competitiveness.  Figure~\ref{GCI} suggests that, in order to become a
rich country, an initially poor country must improve its
competitiveness, but a static approach does not tell us how quickly this
can be achieved. In order to demonstrate the benefit of having a more
competitive economy, we use a ``dynamic'' approach to focus on economic
growth during recession years.  We focus on recession years because
during good years even countries with weak growth policies, e.g.,
countries that use massive indebtedness to increase their GDP, may
experience economic expansion.
  
\begin{figure}[!ht]
\centering   
\includegraphics[width=0.45\textwidth]{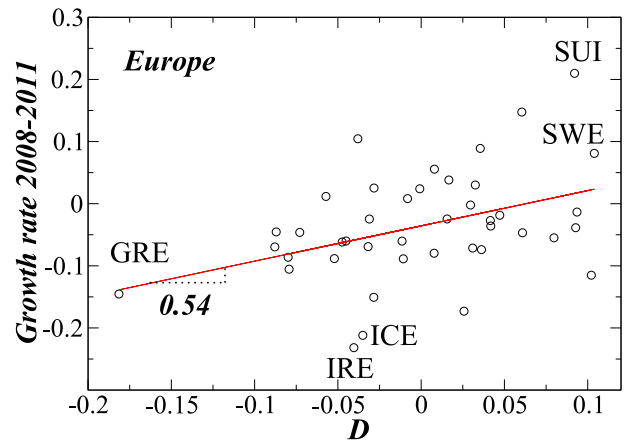} 
\caption{Good to be better than the average during an economic downturn 
2008-2011. The  more positive the relative competitiveness ${\cal D}$ 
 of Eq.~(\ref{D}),  
 the larger  {\it gdp} growth rate for the given period.
 }
\label{DRR}
\end{figure}

In order to test how the level of competitiveness affects the dynamics
of a country's wealth during an economic downturn, we divide 
 the European  countries into two subsets: 
(i) countries with better than average relative competitiveness, for
which ${\cal D}$ defined in Eq.~(\ref{D}) is positive, and (ii)
countries with less than average relative competitiveness, for which
${\cal D}$ is negative. 
 We apply
the statistic for the difference between means and find that during the
2008--2011 period, countries with a positive ${\cal D}$ experienced a
smaller drop in {\it gdp\/} than countries with a negative ${\cal D}$,
where the $t$-statistic yields  2.57 ($d_f = 40$) suggesting that the
difference between these two groups is significant.  
Figure~\ref{DRR} shows the {\it gdp\/} growth rate vs.  ${\cal D}$
defined in Eq.~(\ref{D}) for each of the European countries. Note that it is
clearly advantageous to be better than average.  For the regression line
we obtain a slope of $0.54 \pm 0.20$.  
 For the EU countries, including Croatia, during the
2008--2011 period, 
   the {\it gdp\/} growth rate vs.  ${\cal D}$ yields a slope
   of $0.35 \pm 0.17$.

Addressing question (ii), policymakers need to be able to estimate the
probability that a country will improve its wealth position within the
next 10 years---e.g., will move from ``developing'' to ``developed''
status.  The long-range data show that only a few countries, e.g.,
Singapore and the Republic of Korea, have been able to move from
undeveloped to highly developed during a period of only a few
decades. Such an occurrence can be justifiably classified a rare
event. How probable is it that a poor country such as Peru will become a
rich country like the USA within a decade?  For both academics and
policymakers, quantifying that probability is a problem that deserves
much attention.
    
\begin{figure}[!ht]
\centering  
\includegraphics[width=0.45\textwidth]{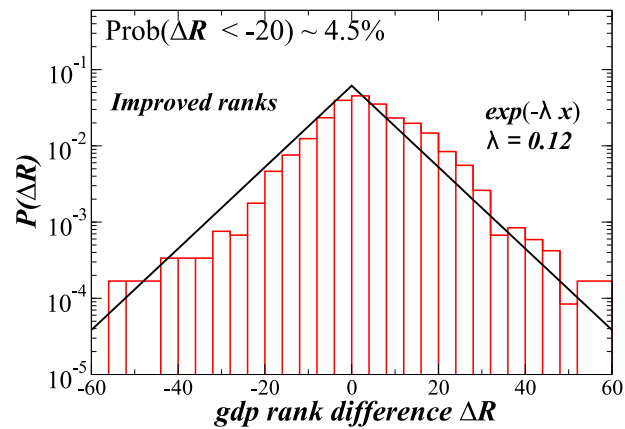} 
\caption{Predictive power of dynamical ranking approach: Probability 
  that a country  increases $(\Delta R < 0)$ or decreases $(\Delta R > 0)$
  its wealth quantified by its {\it per capita}
  GDP ({\it gdp}) rank within a 10-year period.  Pdf  follows an exponential
  distribution obtained by maximum likelihood approach 
   with decay constant 0.12. }
\label{rank}
\end{figure}

To determine this probability we apply Zipf ranking
\cite{Zipf,Stanley95,Gabaix99a,Axtell01,BPPNAS10,BPPNAS11} 
 to {\it gdp\/} over the 32-year
period 1980--2011 \cite{imf}.  For each year $t$ we rank {\it gdp\/} for
an unchanging group of 137 countries from poorest to richest. The
smaller the rank $R_i$, the larger the {\it gdp}, implying that the
decrease in rank, $\Delta R < 0$, corresponds to an improvement in
country wealth.  Using overlapping windows for each initial year $t$ we
calculate the change in rank, $\Delta R$, for each country over a
decade, from $t$ to $t+10$. Figure~\ref{rank} shows that the probability
distribution function (pdf) that a randomly chosen country will increase
its wealth---quantified by its rank---within a decade follows a double
exponential distribution
\begin{equation}
  P(\Delta R) = \lambda / 2 \exp(- \lambda |\Delta R |), 
 \label{exp}
\end{equation}
with decay constant  $\lambda \equiv N/\sum_i |\Delta R_i| = 0.12$ 
 obtained  by applying  a maximum likelihood, 
 which provides useful information
 for policymakers.  Note
 that the left and right tails in Fig.~\ref{rank} correspond to countries
whose wealth improves or worsens, respectively.  To estimate the
probability that Bulgaria's {\it gdp\/} will reach Germany's {\it gdp\/}
in 10 years, we calculate $\Delta R$---the difference between the
current Bulgarian rank and German rank---and, using
Fig.~\ref{rank} and Eq.~(\ref{exp}), we determine the probability $0.5
\exp(- \lambda |\Delta R |)$ that at least a change in rank $\Delta R$ will occur
within 10 years.
 
\begin{figure}[!ht]
\centering  
\includegraphics[width=0.45\textwidth]{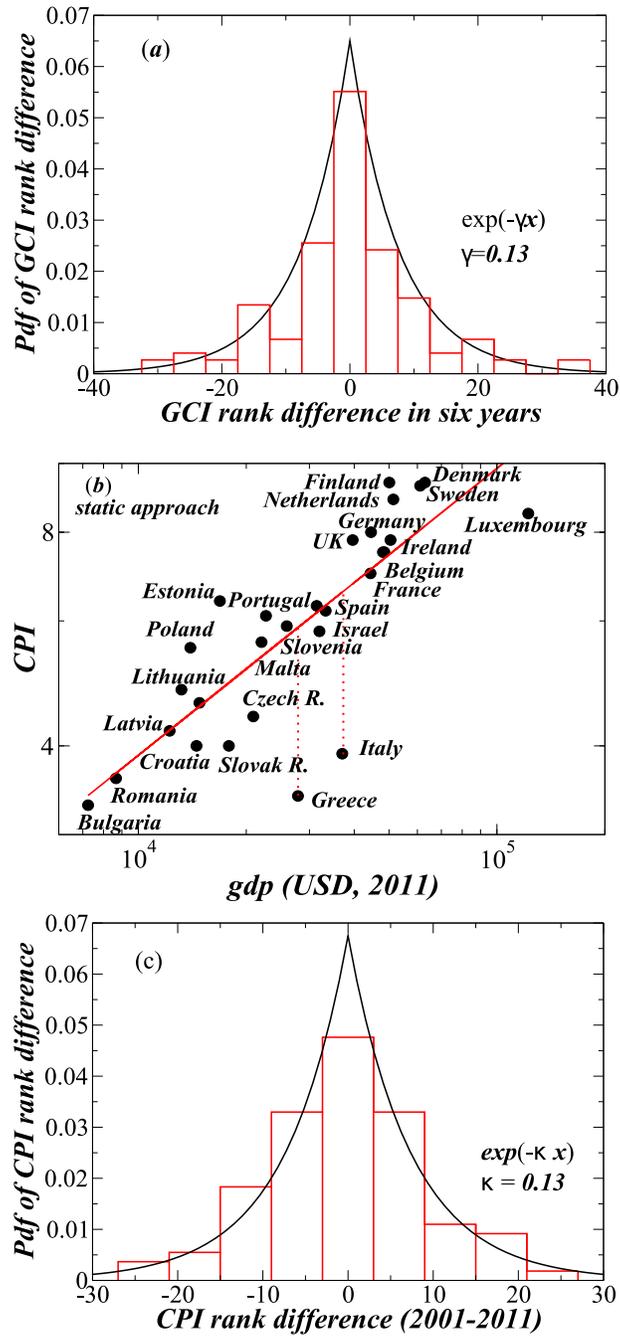}  
\caption{Quantifying the Interplay of the Competitiveness,  
 Corruption,  and Growth of a
  Country. We 
  show (a) a pdf of change in GCI rank between 2005 and 2011 for 124
  countries for which the data we have in this range; 
   (b)   
   corruption relative to country's wealth for 
   EU members including Croatia that will join EU in 2013;    
   (c) pdf of change in CPI
  rank for 91 countries for which the data we have in 2001. We show that
 the decay constant $\gamma$ of  the change in GCI rank and $\kappa$
  of the change in CPI rank are in agreement with the decay constant in 
 Fig.~\ref{rank}.   Pdfs in (a) and (c) approximately 
  follow  exponential
  distributions obtained by maximum likelihood approach. In (b)
   Israel is shown but does not contribute to the fit.  
   }
\label{CPI}
\end{figure}

We begin by assuming that there is a relationship between the
competitiveness and the economic growth of a given country.  To explain
why the probability that the wealth ranking of a country will
substantially change over a decade is low, we study the probability that
GCI values will change over the same period.  Because we lack GCI data
over the 10-year period, we study a pdf of changes in GCI rank for 124
countries during the 6-year period 2005--2011. Although we use GCI data
for a period that differs from the one we used in Fig.~\ref{rank},
Fig.~\ref{CPI}(a) shows the pdf of changes in GCI rank with an exponent
similar to the one in Fig.~\ref{rank}, implying that the dynamics of
{\it gdp\/} and the dynamics of GCI may share the same origin. We
approximate the pdf with an exponential function obtained by using a
maximum likelihood approach.  Because the GCI value for any given
country includes the corruption level of that country, it is reasonable
to assume that the dynamics of the Corruption Perceptions Index (CPI)
\cite{www1,Corruption1,Corruption2} approximates the dynamics of
GCI. Thus for European countries GCI versus CPI follows a power law with
a slope 3.17 ($t$-value 13.1) with a correlation coefficient of 0.90.

Figure~\ref{CPI}(b) shows the 2011 CPI versus the 2011 {\it gdp\/} for
the EU countries.  Because corruption strongly affects a country's
prospects for improvement in its wealth rank over a decade
(Fig.~\ref{rank}), we next study how corruption changes over time.
Figure~\ref{CPI}(c) shows a pdf of changes in CPI rank over the period
2001--2011 for 91 countries.  In agreement with Fig.~\ref{rank}, which
shows the rank of country's wealth, Fig.~\ref{CPI}(c) shows that it is
improbable that a country will rapidly improve its corruption rank.
Again we approximate the pdf of changes in CPI rank with an exponential
function and obtain the exponent $\kappa = 0.13$, similar to the
exponent we found in Fig.~\ref{rank}, implying that the dynamics of {\it
  gdp\/} and the dynamics of CPI may share the same origin.
   	 
Figure~\ref{CPI}(c) shows that during the period 2001--2011 the largest
decreases in CPI rank among the EU countries occurred in Italy (from
rank 29 to 45) and in Greece (from rank 42 to 51).  In Italy the CPI
decreased from 5.2 to 3.9, in Greece from 4.2 to 3.4, in Spain from 7.1
to 6.2, and, surprisingly, in the UK from 8.7 to 7.8.  The largest
increase in CPI occurred in Poland (from 4.0 to 5.5), Estonia (from 5.6
to 6.4), and Germany (from 7.3 to 8.0).  During the last decade, the CPI
values of the majority of the EU countries changed only slightly.  A
change in nation's mentality in this regard appears to be another rare
event.  It is not surprising that the countries with the sharpest drops
in CPI, e.g., Greece and Italy, now face the most dangerous public debt
crises.  This is important information for financial institutions, since
such countries---those with a sharp drop in CPI---become increasingly
risky financially over time.
  
\begin{figure}[!ht]
\centering 
\includegraphics[width=0.45\textwidth]{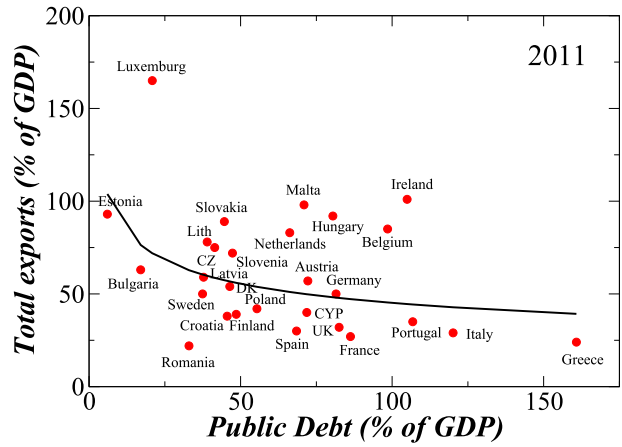} 
\caption{Total export versus public debt, both as a percentage of GDP.
  We fit the plot with a power law and obtain exponent $-0.30 \pm
  0.14$.}
\label{export}
\end{figure}

Finally, to relate indebtedness and competitiveness, Fig.~\ref{export}
shows for 2011 total exports (both goods and services) \cite{worldbank}  
 versus total public debt \cite{eurostat},
both calculated as a percentage relative to a country's GDP. Clearly,
the total export over GDP may be considered another measure of a
nation's competitiveness.  Note that most of the Mediterranean countries
are characterized by a very small total export relative to GDP, which
means that they are generally not very competitive. Some of them, e.g.,
Greece and Italy, are also highly indebted.

\textbf{Modeling political corruption.}
Many factors such as education, technological progress, macroeconomic
stability, good governance, business firm sophistication, and market
efficiency affect productivity and competitiveness
\cite{Schultz,Romer2,Lucas,Fischer,Alesina,martin96,barrosalabook,Solow,Gatti2,Dimatteo1,Aus07,Dimatteo2,BP11,Hidalgo1,Hidalgo2}.
Here we focus on how political corruption affects the growth of a
country's wealth \cite{Pluchino}.

\begin{figure}[!ht]
\centering  
\includegraphics[width=0.3\textwidth]{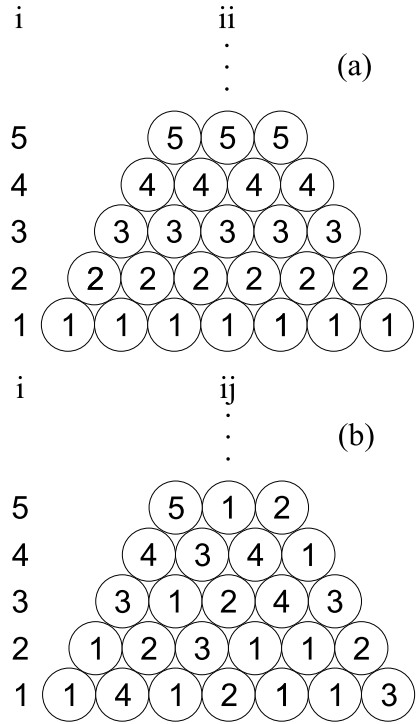} \\ 
\caption{Model cases of political corruption where jobs are
  characterized by different level of skills. (a) Political corruption
  does not exist, since a person with skills $x_i$ occupies an optimal
  position $X_i$, where $x_i = X_i$. In contrast, in (b) political
  corruption exists since generally $x_i \ne X_i$. Job place $X_i$ can
  be occupied by either over-qualified $(x_i > X_i)$ or under-qualified
  $(x_i < X_i)$ person.}
\label{corruption}
\end{figure}

    Many papers have studied 
 collective behaviour of large groups of individuals 
 \cite{Farmer,Lazear,Fairburn,Buchanan,Klimek,Castellano,Helbing}. 
Suppose an imaginary country has only public sector jobs and that there
are only two of them: butchers and neurosurgeons.  Having the butchers
butcher and the neurosurgeons perform neurosurgery is clearly more
efficient than vice versa. In our simplified modeling of political
corruption as an aspect of GCI, each country is of the same size but the
more developed countries have a highly educated and skilled
population. Our model is partially influenced by job-matching models for
the private sector \cite{JovaPE,Miller}.  In our model, each country has
jobs $X_i$ taken from a Poisson distribution that is characterized by a
single parameter $\mu$ where, for example, the USA has a larger $\mu$
than Albania.  In a country with no political corruption [see
  Fig.~\ref{corruption}(a)], the labor force is optimally distributed,
i.e., a worker with skills $x_i$ occupies an optimal job $X_i$, where
$x_i = X_i$, and this holds for every $i$. In a corrupt country [see
  Fig.~\ref{corruption}(b)], people are often given public sector jobs
because of political corruption, nepotism, and bribery, not because they
have the skills. Thus countries with political corruption are not as
efficient in terms of per capita GDP as countries without corruption. 
      	 
To model political corruption and nepotism in the public sector (see
Fig.~\ref{GCI}), we assume that at each time $i$ a new job, $X_i$ is
created, taken from a Poisson distribution, where the larger the Poisson
variable $X_i$, the more skilled the labor required by the position.
Thus job $X_i$ can be held by a worker where $x_i = X_i$ (he/she is
qualified), where $x_i > X_i$ (over-qualified), and also where $x_i <
X_i$ (under-qualified).  We assume that the cases $x_i > X_i$ and $x_i <
X_i$ are less than optimal, i.e., $x_i = X_i$.  It is clear that when
$x_i < X_i$ (the under-qualified case) the worker will not be as
effective as when $x_i = X_i$, but also that when $x_i > X_i$ the worker
will lack motivation and be less efficient.  Specifically, we assume
that each worker $x_i$ in the public sector works at a job $X_i$ where
there is a discrepancy $(x_i \ne X _i)$ controlled by a Gaussian
distribution centered at $X_i$ with a standard deviation $\sigma$. This
choice of mean $(X_i)$ allows that, when $\sigma=0$ and the country has
no corruption, for each $i$, $x_i=X_i$ (see Fig.~\ref{corruption}), and
workers will be optimally distributed to jobs.  As $\sigma$ increases,
political corruption increases.  For each $i$, there is a discrepancy
given by
\begin{equation} 
  \delta_i = \exp(- |x_i - X_i|),   
\label{ei}
\end{equation} 
and this functional dependence value allows $\delta_i ~\epsilon ~
[0,1]$. Utilizing Ref.~\cite{Axtell08,Pluchino}, the total discrepancy in the
group with $N$ agents is
\begin{equation} 
  E = \Sigma_{i=1}^{N}  \delta_i.
\label{E}
\end{equation}   

The entire group produces output (GDP) as a function of $E$, where we
assume
\begin{equation} 
 {\rm GDP} =  \mu ~E, 
\label{GDP}
\end{equation}   
where ${\it gdp}$ = GDP$/N$ and $\mu$ comes from the Poisson
distribution of jobs $X_i$ introduced above.  Note that for a country
with no corruption---$\sigma=0$, implying $\delta_i = 0$ for each
$i$---Eqs.~(\ref{ei}) and (\ref{E}) yield $E=N$, and Eq.~(\ref{GDP})
yields maximum GDP.  We introduce $\mu$ in Eq.~(\ref{GDP}) because if we
compare a country with no corruption but with badly educated and
unskilled citizens with $\mu_1$ with a country with no corruption but
with very educated and highly skilled citizens with $\mu_2$, where
$\mu_1 < \mu_2$, it must hold that ${\rm GDP}_1 < {\rm GDP}_2$.
 Note that Eq.~\ref{GDP} in our framework relates country wealth, 
 education level,
 and corruption level, quantified by GDP, $\mu$, and $E$, respectively.
	
\begin{figure}[!ht]
\centering  
\includegraphics[width=0.4\textwidth]{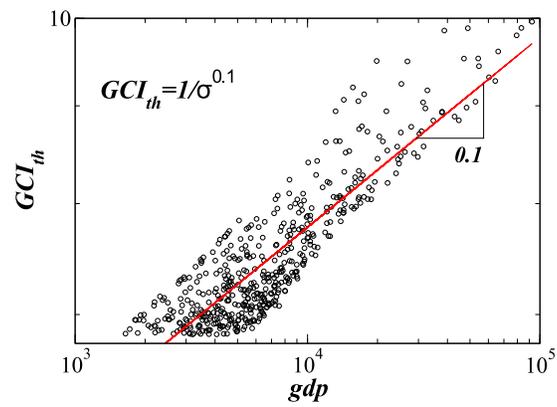} 
\caption{Modeling how well a society is organized quantified by
  parameter $\sigma$.  Model simulations for ${\rm GCI}_{th} $ vs. model
  GDP.  We define theoretical CPI as ${\rm GCI}_{th} \equiv 1/ \sigma
  ^{\gamma}$.  By varying $\gamma $ we change the slope between ${\rm
    GCI}_{th} $ and {\it gdp}.  }
\label{model1}
\end{figure}

We assume that different countries will differ in skill levels $\mu$
according to a Poisson distribution and in corruption levels quantified
by $\sigma$ of a Gaussian distribution.  Figure~\ref{model1} shows 1000
countries each with 10,000 jobs where $\mu$ ranges from 5 to 20, and
$\sigma$ from 0.5 to 20, both taken from uniform distributions.  Note
that GCI is defined such that, as it increases, corruption in a given
country decreases \cite{Schwab}.  We define a theoretical GCI as ${\rm
  GCI}_{th} = \sigma^{-\gamma}$.  With this expression in which $\gamma
= 0.1$, we obtain a power-law dependence between ${\rm GCI}_{th}$ and
{\it gdp}, similar to the one obtained in Fig.~\ref{GCI}.  As ${\rm
  GCI}_{th}$ increases, {\it gdp} also increases.  Clearly, by tuning
parameter $\gamma$ in ${\rm GCI}_{th} = 1/ \sigma ^{\gamma}$, we can
obtain any slope between ${\rm GCI}_{th}$ and {\it gdp} we want.

Note that in order to model Fig.~\ref{rank}, we should assume that the $\mu$
and $\sigma$ of different countries change over time.  Also note that
here we have modeled only the public sector and its contribution to
GDP, but we could have easily included the private sector in our
analysis. For example, Jovanovic \cite{Jovanovic82} has proposed a theory
of selection in which the key is firm efficiency and how it affects firm
growth.
	 
\section{Discussion}
     
What can policymakers do to substantially increase their country's
wealth?  It was reported that institutional integrity strongly affects
competitiveness and growth \cite{Acemoglu01}. In addition, government
attitudes toward markets and freedoms and the efficiency of its
operations are also very important: excessive bureaucracy,
over-regulation, corruption, dishonesty in dealing with public contracts,
and political dependence of the judicial system impose significant
economic costs to businesses and slow the process of economic
development \cite{Schwab}. Potential investors are also unwilling to
invest if their rights as owners are not properly protected.
Furthermore, the extent of centralization of a given market in both the
public and private sectors greatly impacts the competitiveness and
economic development within a given country. Thus a highly centralized
private market can be as problematic as a highly centralized public
sector.

Figure~\ref{rank} shows that it is highly improbable that a country
will experience a substantial increase in its wealth. There are many
reasons for this \cite{Preis}.  One of the most important is that to
experience a substantial increase in wealth a nation must change its
collective behavior. It must, for example, reduce its levels of nepotism
and corruption [see Fig.~\ref{CPI}(b)]. We might imagine a country in
which all citizens are equally educated and equally skilled. In such a
case, political corruption would not affect growth.  In real-world
corrupt countries, however, we can assume that the less-skilled job
holders are more politically connected than the more-skilled job
holders.  Political corruption and nepotism can also account for the
difference in effectiveness between public and private sectors, because
in  corrupt countries many of the public sector jobs are held by
those who are politically connected, irrespective of their
qualifications---a majority of
political party members hold public sector jobs they would not have if
they were not party members. 
  An ineffective public sector often generates an
increase in public debt and, as the public debt level increases, the
government must consider raising taxes. Raising taxes then can
substantially affect the economy's competitiveness. Thus a country's
level of corruption can determine how the country is organized and how
efficient its government is, which, in turn, affects the private sector
that pays the government's bills.

Interdependent groups of countries such as the EU seem to be more
vulnerable to financial fluctuations than independent single countries,
a finding that is agreement with recent studies of interdependent
networks \cite{Buldyrev,Vespignani10,Gao12,Dror1,Dror2}.  Recently, in
order to increase the competitiveness of the EU and reduce its financial
vulnerability, many EU politicians have been advocating not just a
continuation of the currency union, but also the creation of a fiscal
union. Before this is attempted, a common mechanism to control
corruption at the EU level is needed, e.g., an anti-nepotism law. If
corruption is allowed to continue and grow, un-corrupt EU countries will
increasingly be paying the bills of the corrupt.  Fighting corruption is
thus fighting for an increase in competitiveness, and this must be a top
EU priority.
   
\section{Methods}
 
The World Economic Forum has released the Global Competitiveness Index
(GCI), developed by Sala-i-Martin and Artad, in order to assess the
ability of countries to provide high levels of prosperity for their
citizens \cite{Schwab}. The index quantifies how productive a country is
as it uses available resources.  It measures the microeconomic and
macroeconomic foundations of national competitiveness. GCI is defined as
a weighted average of many different components, each measuring a
different aspect of competitiveness: institutions, infrastructure,
macroeconomic environment, health, primary education, goods market
efficiency, labor market efficiency, financial market development,
technological readiness (e.g., access to high technology), market size,
business sophistication, and innovation \cite{Schwab}.  It provides a
raw score that ranges between 0 and 6, where the later value defines the
most competitive country.  The 2011--2012 Global Competitiveness Report
covers 142 developed and developing economies.

We also analyze the Corruption Perceptions Index (CPI)
\cite{www1,Corruption1} introduced by Transparency International.  The
CPI is a composite index based on independent surveys of business people
and on assessments of corruption in different countries provided by more
than ten independent institutions around the world, including the World
Economic Forum, the United Nations Economic Commission for Africa, and
the Economist Intelligence Unit. The CPI ranges from 0 (highly corrupt)
to 10 (highly transparent).
 
\subsection*{Acknowledgments.~~}

We thank the ONR and the  Keck Foundation for financial support.

\subsection*{Author contributions.~~}

BP, DH, and DYK performed analyses, and BP, DH, DYK and HES discussed
the results, and contributed to the text of the manuscript.

\subsection*{Additional Information.~~}
The authors declare no competing financial interests.

\newpage

\end{document}